\documentclass[preprint,12pt]{elsarticle}

\usepackage{amssymb}
\usepackage{bm} 
\usepackage{graphicx}

\newcommand{\Pomeron}{I\!\!P}

\usepackage{epstopdf}
\journal{Nuclear Physics B}

\begin{document}

\begin{frontmatter}



\title{Leading twist shadowing,  black disk regime and
 forward hadron production}


\author{Mark Strikman}
\address{104 Davey Lab, Penn State University, University Park, PA 16803, U.S.A.}

\begin{abstract}
We review theory of the leading twist nuclear shadowing, and  describe phenomenon of post-selection suppression of leading parton spectrum (effective fractional energy losses) in the proximity of the black disk regime. We argue that
$2 \to 2$ mechanism dominates in the inclusive leading pion production 
in d-Au collisions and explain that 
 the post-selection naturally explains both the magnitude of the suppression of the forward pion production in d-Au collisions and the pattern of the forward - central correlations. At the same time this pattern of correlations   rules out $2\to 1$ mechanism as the main source of the inclusive leading pion yield. It is demonstrated that the mechanism of the  double parton interactions gives an important contribution to the production of two leading pions in $pp$ scattering opening a new way to study correlations of leading quarks in the nucleon. The same mechanism is enhanced in $dAu \to \pi^0\pi^0 +X$ collisions and  explains the dominance of $\Delta\varphi$ independent component and suppression of the away side peak. 
\end{abstract}


\end{frontmatter}


\section{Theory of the leading twist nuclear shadowing}
\label{shadowing}
The connection between nuclear shadowing and diffraction was established a long time ago \cite{Gribov:1968jf}. One can understand it as a manifestation of the unitarity as reflected in the Abramovsky, Gribov, Kancheli (AGK) cutting rules \cite{Abramovsky:1973fm}. Accuracy of the theory for the hadron - nucleus interactions is on the level of few \% which reflects small admixture of non-nucleonic degrees of freedom in nuclei and small off-shellness of the nucleons in nuclei as compared to the soft strong interaction scale. 
The Gribov logic  was successfully applied for the description of the $\gamma A$ total cross sections. However an additional  step is necessary  to calculate the nuclear parton distributions at small $x$ \cite{Frankfurt:1998ym}. One has to combine 
unitarity relations for different cuts of the diagrams corresponding to the different final states for the interaction of the hard probe with a nucleus, with the QCD factorization theorem for hard diffraction \cite{Collins:1997sr}. In the limit of thin target when only double scattering is important one obtains
\begin{eqnarray}
\Delta \left[ x f_{j/A}(x,Q^2,b)\right]=  x f_{j/N}(x,Q^2,b) -x f_{j/A}(x,Q^2,b)  & & \nonumber \\
=  8 \pi A(A-1) \Re e \left(\frac{(1-i\eta)^2}{1+\eta^2}\right)  \int^{0.1}_x d x_{\Pomeron}
\beta f_j^{D(4)}(\beta,Q^2,x_{\Pomeron},t_{{\rm min}})& & \nonumber \\ 
\times \int^{\infty}_{-\infty}d z_1  \int^{\infty}_{z_1}  d z_2 \, 
\rho_A(\vec{b},z_1) \rho_A(\vec{b},z_2) 
e^{i (z_1-z_2) x_{\Pomeron} m_N}, & &
\label{eq:m12}
\end{eqnarray}
where $ f_{j/A}(x,Q^2), f_{j/N}(x,Q^2)$ are nuclear and nucleon pdfs, $ f_j^{D(4)}(\beta,Q^2,x_{\Pomeron},t_{{\rm min}})$ are diffractive pdfs, $\eta =Re A^{diff}/ Im A^{diff}\approx 0.17$ and $  \rho_A(r)$ is the nuclear matter density.

Numerical studies indicate that dominant contribution to the shadowing comes from the region of relatively $\beta=Q^2/(M^2+Q^2)$ corresponding to rapidity intervals $\le 3$ for which small $x$ approximation used in the BFKL approaches is not applicable (these approaches lead also  to the diffractive cross section with $\alpha_{\Pomeron} \sim 1.25$ while the HERA experiments find a  soft energy dependence corresponding to the effective Pomeron trajectory with $\alpha_{\Pomeron} \sim 1.11$).

\begin{figure}[h]  
   \centering
   \includegraphics[width=0.9\textwidth]{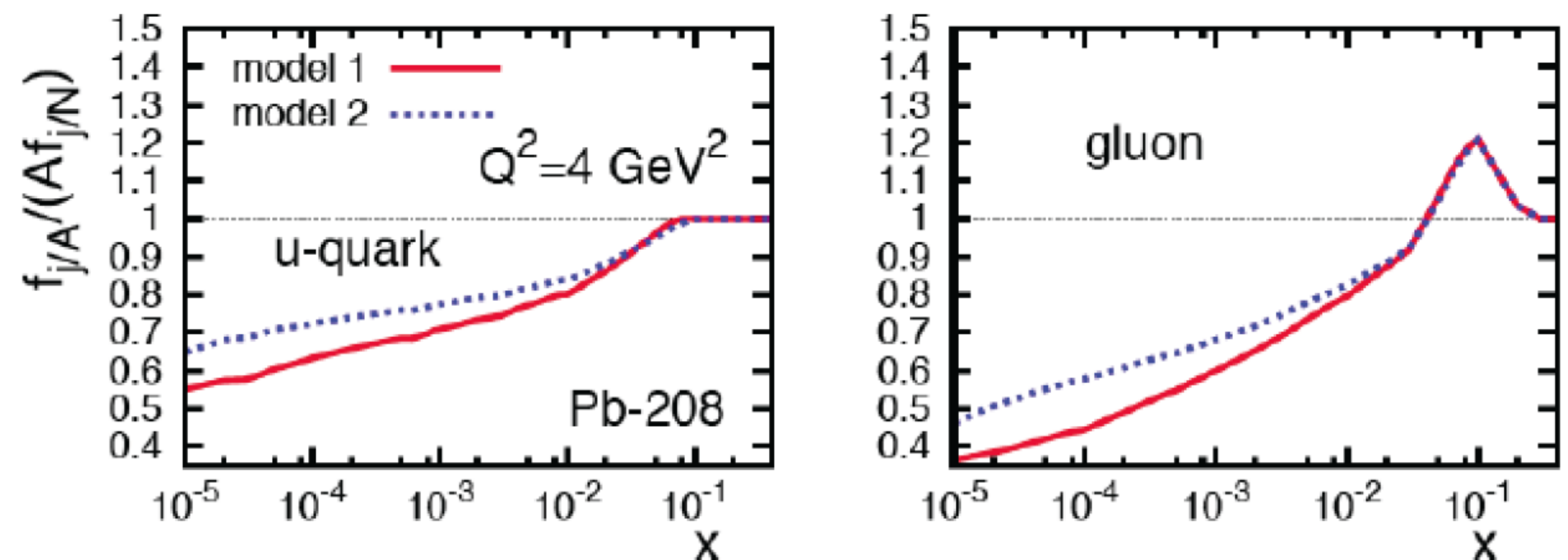} 
   \caption{Predictions of the theory of leading twist shadowing for quark and gluon pdfs. Difference in the predictions of two models of color fluctuation illustrates the range of uncertainties of the predictions.}
    \label{shadfig}
 \end{figure}

 The uncertainties of the predictions are related to the shadowing effects resulting from the interaction of the hard probe with $N \ge 3$ nucleons. 
 Recently we improved the
  treatment of the  multiple interactions,  based on the concept of the color fluctuations and accounts for the presence of both point-like and hadron-like configurations in the virtual photon wave function \cite{Guzey:2009jr}.
This allowed us  to reduce significantly the uncertainties: the difference between two extreme scenarios (models I and II in Fig.\ref{shadfig}) is  $\le  20\% $ for $A\sim 200$ and much smaller for light nuclei. 

\section{Where a non-linear regime sets in?}
 
 In the leading log approximation one can derive a relation between the QCD evolution equations and the target  rest frame picture of the interaction of 
 small  color dipoles with targets expressing it through the gluon density in the target, see  \cite{Frankfurt:1996ri}
and refs. therein. Matching the behavior of the dipole cross section in the pQCD regime and of large size dipoles in the regime of soft interaction,  it is possible to write interpolation formulae for the dipole - nucleon cross section for all dipole sizes and describe
 the total cross section of DIS at HERA. 
 
 To determine how close is the interaction strength to the maximal allowed by the unitarity it is necessary to consider the amplitude of the dipole - nucleon interaction in the impact parameter space:
 \begin{equation}
 \Gamma_h(s,b)={1\over 2is}{1\over (2\pi)^2} \int d^2\vec{q}e^{i\vec{q}\vec{b}}A_{hN}(s,t),
 \end{equation}
 where $ A_{hN}(s,t)$ is the elastic amplitude of dipole - nucleon scattering normalized to $\it{Im} A_{hN}= is_{hN}  \sigma_{tot}(hN)$. The limit $ \Gamma_h(s,b)=1$ corresponds to the regime of the complete absorption - black disk regime - the maximal strength allowed by the S-channel unitarity.
  
 The   $t$ dependence of the dipole - nucleon elastic scattering amplitude can be obtained from the studies of the exclusive vector meson production in the regime where QCD factorization theorem for the exclusive processes allows to express relate the  $t$ dependence of the amplitude to the $t$-dependence of the gluon GPDs. 

\begin{figure}[h]  
   \centering
   \includegraphics[width=0.9\textwidth]{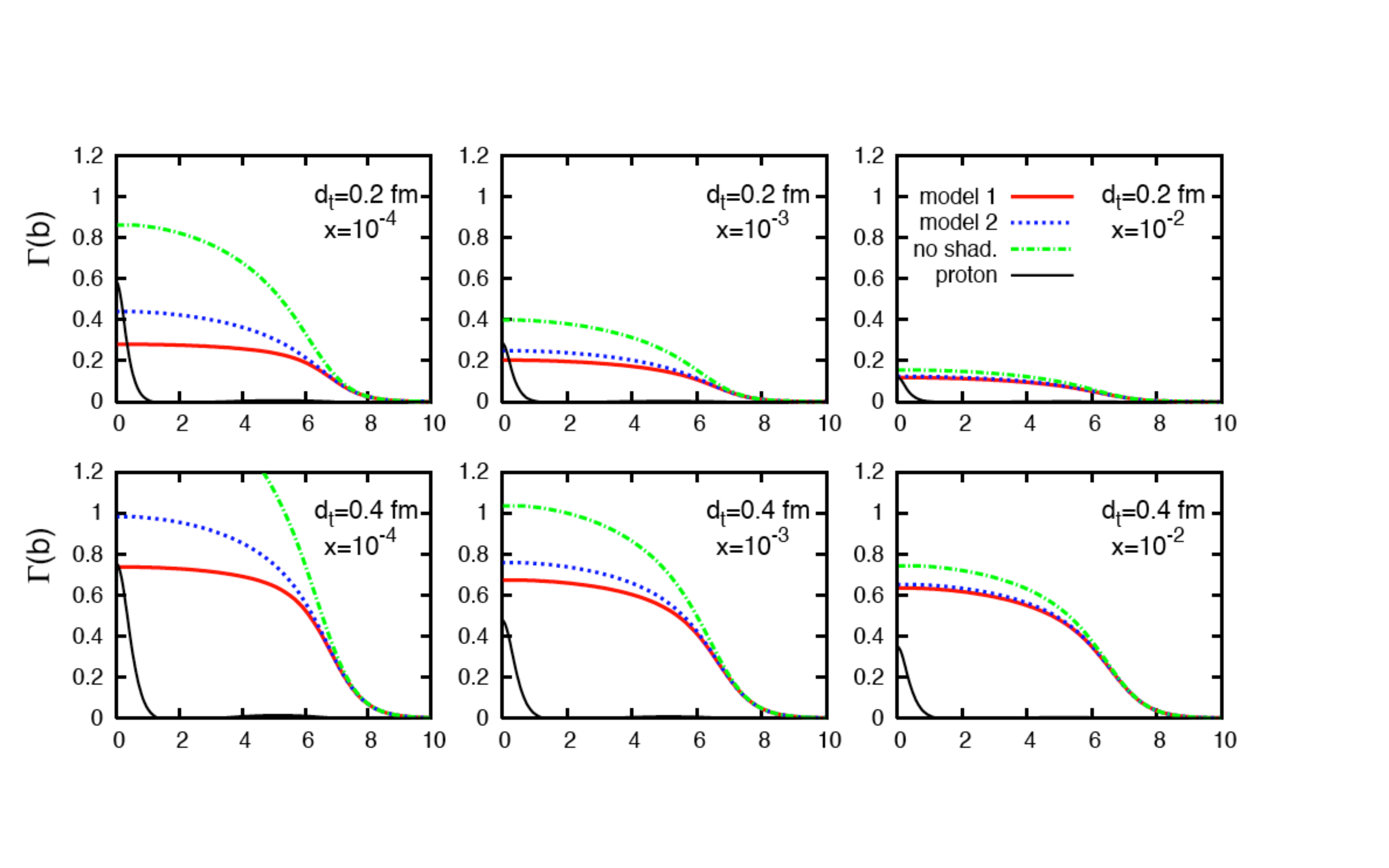} 
   \vspace{-1.0cm}
   \caption{Impact parameter distribution of $q\bar q $ dipole interaction with protons and lead nucleus. The models for the gluon shadowing are the same as in Fig.\ref{shadfig}.}
    \label{impactfig}
 \end{figure}

Combining this information with the information on the total cross section of the dipole - nucleon interaction allows us to determine  $ \Gamma_h(s,b)$  as function of the dipole size.  A sample of the results for $q\bar q$ dipole -proton (nucleus) interaction which represent an update of the analysis of \cite{Rogers:2003vi}
is presented in Fig.\ref{impactfig}. One can see from the figure that for small impact parameters $\Gamma$ for scattering off a proton and a heavy nucleus are comparable  provided one takes into account the leading twist shadowing for the gluons.
This reflects an observation that gluon densities in nuclei and proton at $b=0$ are rather  similar.  However very few processes with proton are sensitive to $b=0$  rather than  $\left<b\right>$,  while in nuclei   $\Gamma(b)$ is practically constant
for a broad range of $b$.

\vspace{-0.5cm}

Since the  probability of the inelastic interaction,  $P_{in}(b)= 
1 -\left|1-\Gamma(b)\right|^2$,    $P_{in}\ge 3/4 $ already for $\Gamma(b)\ge 0.5$. Also, in the case of the gluon induced processes $P_{in}(b)$ is enhanced by a factor 9/4 (provided it is sufficiently small). Hence  the results presented  in Fig.\ref{impactfig} indicate
that gluon induced  interactions are close to the BDR for a much larger range of the dipole sizes.

 \section{Post selection effect in BDR - effective energy losses}
 It was argued in \cite{Frankfurt:2001nt} that in the BD regime  interactions with the target select configurations in the projectile wave function where  the projectile's energy is split between constituents much more efficiently  than in the DGLAP regime. The simplest   example is inclusive production of the leading hadrons in DIS for $Q\le 2 p_t(BDR)$. Interactions with the target are not suppressed up to   $p_t\sim p_t(BDR)$, leading to selection of configurations in $\gamma^*$ where longitudinal fractions carried by quark and antiquark are comparable. The energy splits before the collision - post-selection. As a result to a first approximation the leading hadrons are produced in the independent fragmentation of $q$ and $\bar q$:
\begin{eqnarray}
\bar{D}^{\gamma^{\ast}_{T}\to h}(z)=2 \int^{1}_{z}dy D^{h}_{q}(z/y) \frac{3}{4}(1+(2y-1)^2),
\end{eqnarray}
leading to a strong suppression of the  hadron production at $x_F> 0.3$.

In the case of a parton  of a hadron  projectile propagating through the nucleus 
near BDR effective energy losses were estimated in Ref.\cite{Frankfurt:2007rn}. For quarks they  are expected to be  of the order of 10$\div$ 15  \% in the regime of the onset of BDR and larger deep inside this regime. Also the effective energy losses  are somewhat larger for gluons as the $g\to$ gg splitting is more symmetric in the light cone fractions than the
 qg splitting.

\section{Leading hadron production in hadron - nucleus scattering}
\label{single}
Production of leading hadrons with $p_t\sim \mbox{few \, GeV/c}$ in hadron - nucleus scattering at high energies provides a sensitive test of the  onset of the BDR dynamics. Indeed in this limit pQCD provides a good description the forward single inclusive pion production in $pp$  scattering \cite{Werner}. At the same time it was found to overestimate grossly   the cross section of the pion production in $dAu$ collisions in the same kinematics. The analysis of ~\cite{GSV} has demonstrated that the dominant mechanism of the single pion production in the $NN$ collisions in the kinematics which was studied at RHIC is scattering of leading quark of the nucleon  off the gluon of the target
with the median value of $x$ for the gluon to be in the range $x_g \sim 0.01 \div 0.03$ depending on the rapidity of the pion.  The nuclear gluon density for such x is known to be close to the incoherent sum of the gluon fields of the individual nucleons since the coherent length in the interaction is rather modest for such distances (cf. Fig.\ref{shadfig}). As a result the leading twist  nuclear shadowing effects can explain only a very small fraction of  the observed suppression \cite{GSV} and one needs a novel dynamical  mechanism to suppress generation of pions in such collisions. It was pointed out in \cite{GSV} that the energy fractional 
energy losses on the scale of $10\div 15\%$ give a correct magnitude of suppression of the inclusive spectrum due to a steep fall of the cross section with $x_F$ which is consistent with the estimates within the 
 post-selection mechanism.
 
 An important additional information comes from the correlations studies where correlation of the leading pion with the pion produced at the central rapidities x 
 \cite{star,phenix2} which corresponds to the kinematics which receives   the dominant contribution from the scattering off gluons with $x_g\sim 0.01 \div 0.02$. The rate of the correlations for $pp$ scattering is consistent with pQCD expectations.
An extensive analysis performed in \cite{Frankfurt:2007rn} has demonstrated that the strengths of "hard forward pion"  -- "hard $\eta \sim 0$ pion"  correlations in 
$ dAu$ and in $pp$ scattering are similar. A  rather small difference in the pedestal originates from the multiple soft collisions. Smallness of the increase of the soft pedestal as compared to $pp$ collisions unambiguously  demonstrates that the dominant source of the leading pions is the $dAu$  scattering at large impact parameters. This conclusion is supported by the experimental observation \cite{Rakness} that the associated multiplicity of soft hadrons in events with forward pion is a factor of two smaller than in the minimal bias $dAu$ events. A factor of two  reduction 
 factor is consistent with the estimate of  \cite{Frankfurt:2007rn} based on the analysis of the soft component of $\eta =0$ production for the forward pion trigger.
 Overall these data indicate that 
 (i) the dominant source of the forward pion production 
 is $2\to 2$ mechanism, (ii) production is dominated by projectile scattering at large impact parameters, (iii) proportion of small $x_g$ contribution in the inclusive rate is approximately the same for $pp$ and $dAu$ collisions. 
 
 A lack of additional suppression of the $x_g \sim 0.01$ contribution to the double inclusive spectrum as compared to the suppression of the inclusive spectrum is explained in the post-selection mechanism as  due to relatively small momentum of the gluon in the nucleus rest frame putting it far away from the BDR.
 
 It is difficult to reconcile  enumerated  features of the forward pion production data  with the 
 $2\to 1 $ mechanism \cite{Kharzeev} inspired by the color glass condensate model. In the scenario of  \cite{Kharzeev}  incoherent $2\to 2$ mechanism is neglected, a strong suppression of the recoil pion production is predicted. Also it   leads to a dominance of the central impact parameters and hence a larger multiplicity for the central hadron production in the events with the forward pion trigger. The observed experimental pattern indicates the models \cite{Dumitru:2005gt} which neglect contribution of the $2\to 2$ mechanism and consider only  $2\to 1$  processes strongly overestimates  contribution of the $2\to 1$ mechanism to the inclusive cross section.

 \section{Production of two forward pions and double-parton mechanism in $pp$ and $dA$ scattering}

 In Ref. \cite{GSV} we suggested that in order to study the effects of small $x$ gluon fields in the initial state one should   study production of two leading pions in nucleon - nucleus collisions. Recently the data were taken on production of two forward pions in dAu. The preliminary results of the studies of the reactions
 $pp \to \pi^0\pi^0+X,  d-Au \to \pi^0\pi^0+X$,  where one leading  pion served as a trigger and the second leading pion had somewhat smaller longitudinal and transverse momenta \cite{starqm09,phenixqm09}.
  The data indicate a strong suppression of the back to back production of pions in the central $dAu$ collisions. Also  a large fraction   of the double inclusive cross section  is isotropic in the azimuthal angle $\Delta\varphi$ 
of the two pions.  
  In order to understand the origin of the suppression and other features of the data  we performed a study \cite{SV} which is summarized below.
  
  \subsection{Forward dipion production in $pp$ scattering}
  it is instructive to start with the case of $pp$ scattering.
  \begin{figure}[t]  
   \centering
   \includegraphics[width=1.0\textwidth]{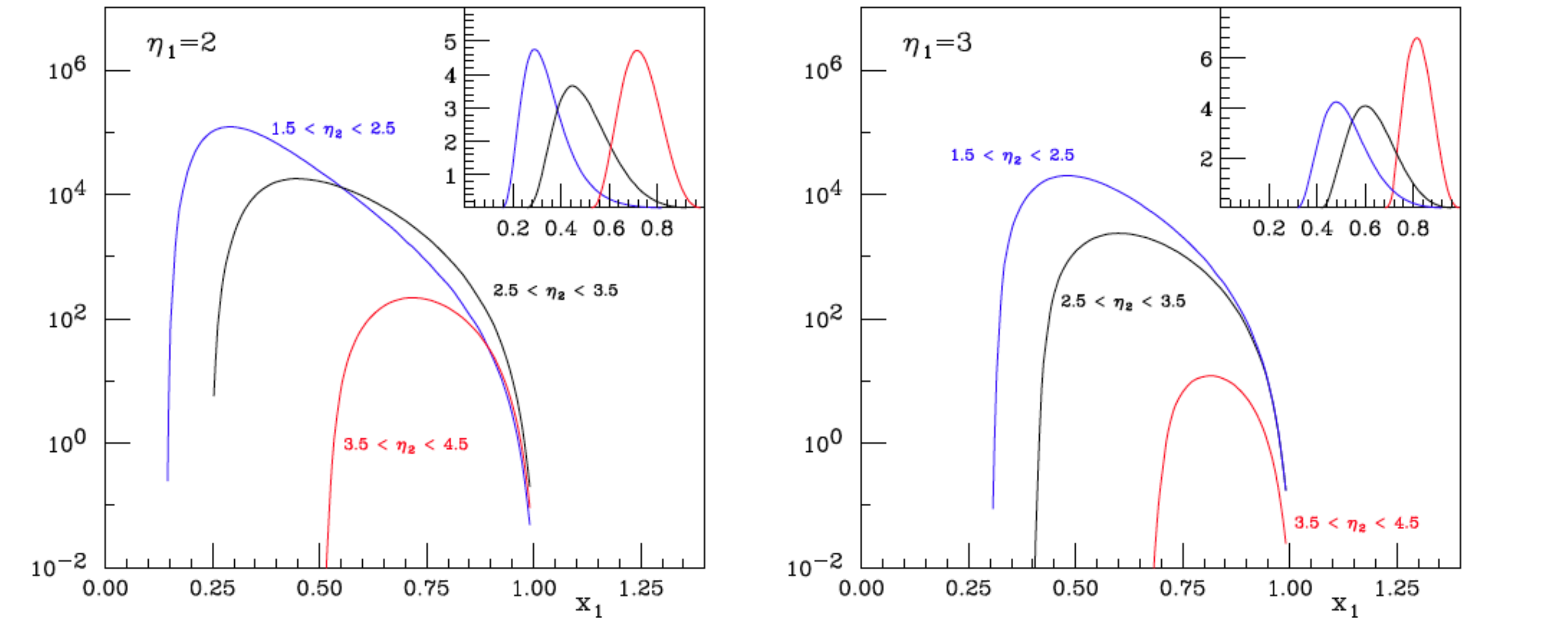} 
 \caption{{ Distributions of the leading-power LO cross section
for $pp\to \pi^0\pi^0X$   at $\sqrt{S}=200$~GeV 
in momentum fraction $x_a$, for $\eta_1=2$ (left) and $\eta_1=3$ (right).
We have chosen $p_{T,1}=2.5$~GeV and integrated over 
$1.5~\mathrm{GeV}\leq p_{T,2}\leq p_{T,1}$ and various bins in $\eta_2$. 
Units are arbitrary. The inserts show the corresponding normalized
distributions on a linear scale.}}
 \label{xdistrfig}
 \end{figure}
  \subsection{Kinematics of the $2\to 2$ mechanism of two pion production}
We start with the leading twist mechanism $2\to 2$ in which a leading quark from the nucleon and a small x gluon scatter to produce two jets with leading pions.  In this kinematics $x_g \le M^{2}(\pi \pi)/ x_q s_{NN}$. Production of two pions which together carry a large fraction of the nucleon momentum can occur only if $x_q$ is sufficiently close to one. 
The cross section is given by the standard expression for the $2\to 2$ processes.
In Fig.\ref{xdistrfig} the results of the calculation of the integrand of the integral over $x_q$ for the LT  mechanism are presented. They show that   average value of $x_q$ for typical cuts of the RHIC experiments is very high.

\subsection{Double parton mechanism}
It is more likely that two rather than one quark in a nucleon carry together $x $ close to one. This suggests that in the   discussed  RHIC kinematics production the "double-scattering" contribution with two separate hard interactions in a single $pp$ collision could become important. Hence though  the discussed contribution is a "higher-twist", it is enhanced both by the probability of the relevant two quark configurations and the increase of the gluon density at small x which enters in the double-scattering in the second power.

\begin{figure}[t]  
    \includegraphics[width=1.0\textwidth]{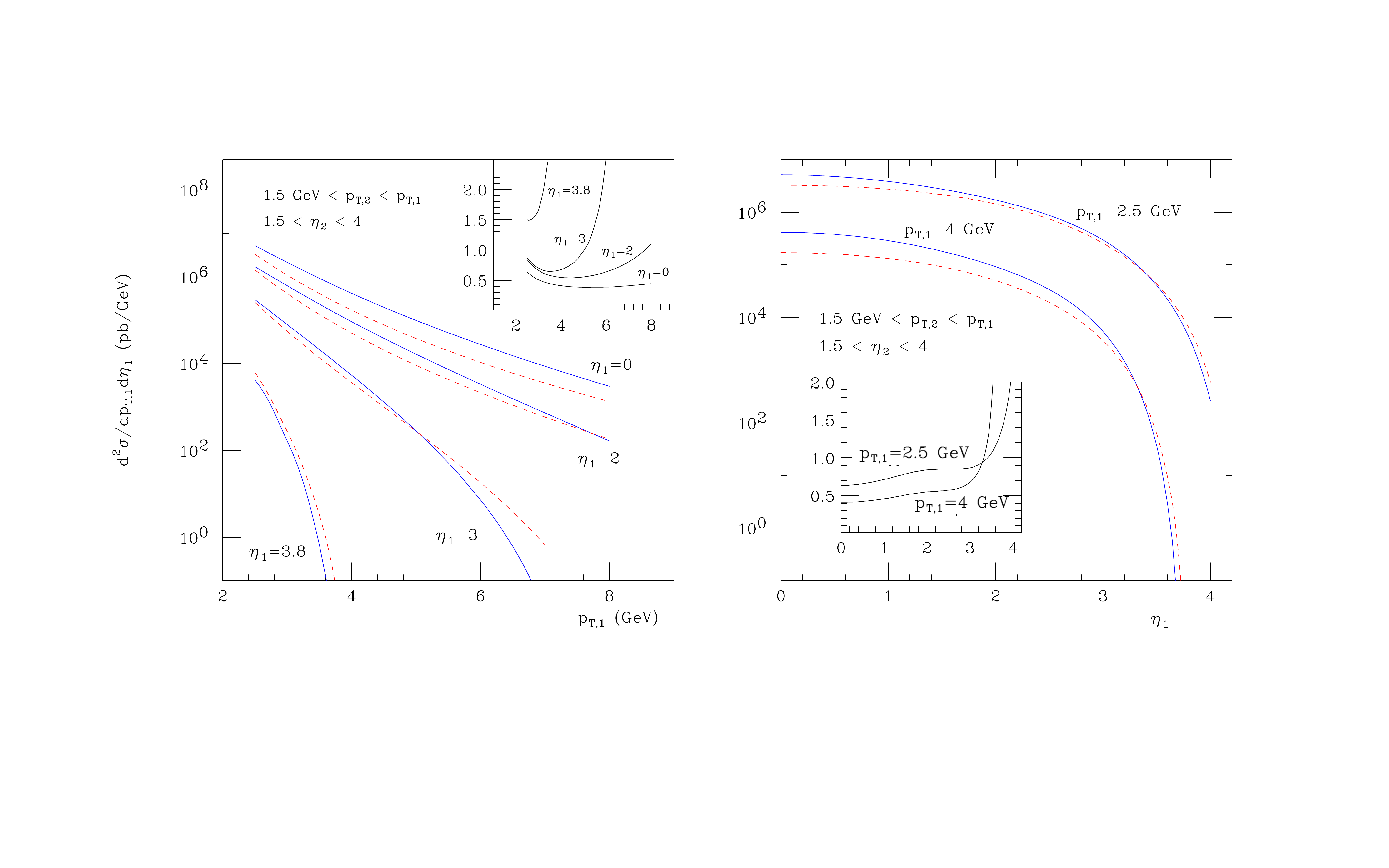}
  \vspace*{-20mm}
   \caption{Comparison of the cross sections of LT  (solid curves) and double parton mechanisms (dashed curves). Inserts show the ratios of double parton and LT mechanisms.}
    \label{MIfig}
 \end{figure}

One can derive the expression for the double-scattering  mechanism based on the analysis of the corresponding Feynman diagrams and express it through 
 the new  rank two generalized parton densities (GPDs)  in the nucleons, see  \cite{BDFS} and refs therein.
 The cross section can be written in the form 
 \begin{eqnarray}
\frac{d^4\sigma}{dp_{T,1}d\eta_1dp_{T,2}d\eta_2} ={1\over \pi R^2_{int}}
\sum_{abcda'b'c'd'}
\int dx_a dx_b dz_c dx_{a'} dx_{b'} dz_{c'} \,\nonumber 
& & \\
 f_{aa'}^{H_1}(x_a, x_{a'})f_b^{H_2}(x_b)f_{b'}^{H_2}(x_{b'})D_c^{h_1}(z_c)\,
D_{c'}^{h_1}(z_{c'})\,
\frac{d^2 \hat{\sigma}^{ab\to cd}}{dp_{T,1} d\eta_1}\,
\frac{d^2 \hat{\sigma}^{a'b'\to c'd'}}{dp_{T,2} d\eta_2}\;
\label{mastermp}
\end{eqnarray}
Here  $f_{aa'}^{H_1}(x_a, x_{a'})$ is the double parton distribution. If the partons are not correlated, it is equal to the product of the single parton distributions.
For simplicity we neglected here correlations in the target as in our case $x's$ for gluons  are small. The  dimensional factor $ \pi R_{int}^2$
 characterizes  the transverse spread of the parton distributions in two nucleons as well as possible transverse parton - parton correlations and can be expressed through the integral of the product of two rank two GPDs (in many experimental papers and some theoretical papers this quantity  is denoted as $\sigma_{eff}$ though it has little to do with the interaction cross section of the colliding hadrons). In the approximation when partons are not correlated in the transverse plane one can express  $\pi R_{int}^2$
through the single parton GPDs, leading to $\pi R_{int}^2 \approx 34 mb$ 
\cite{Frankfurt:2003td}. This is larger than  $\pi R_{int}^2 \approx 15 $ mb observed at the Tevatron, indicating presence of the correlations in the nucleon. In our numerical calculations we will use this experimental value.
  We find that for the RHIC kinematics the only trivial correlation due to the fixed number of the valence quarks is important while the correlation between $x_a$ and $x_{a'}$ 	remains a small correction if we follow the quark counting rules to estimate the $x_{a'}$  dependence off $f_{a,a'}(x_{a},x_{a'}) $ for fixed $x_a$.
 The results of our calculation are presented in Fig.~\ref{MIfig}. They  indicate that 
 that the LT and double parton mechanisms are comparable for the kinematics of the RHIC experiments. This provides a natural explanation for the presence of a large component in the $pp \to \pi^0\pi^0+X$ cross section measured in 
 \cite{starqm09,phenixqm09} which does not depend on the azimuth angle between two pions, 
 $\Delta\varphi$. In fact the   number of events in the pedestal comparable to the peak around $\Delta\varphi \sim \pi$  which is dominated by the LT contribution indicating that the LT and double-parton contributions are comparable.

\begin{figure}[t]  
   \centering
   \includegraphics[width=0.8\textwidth]{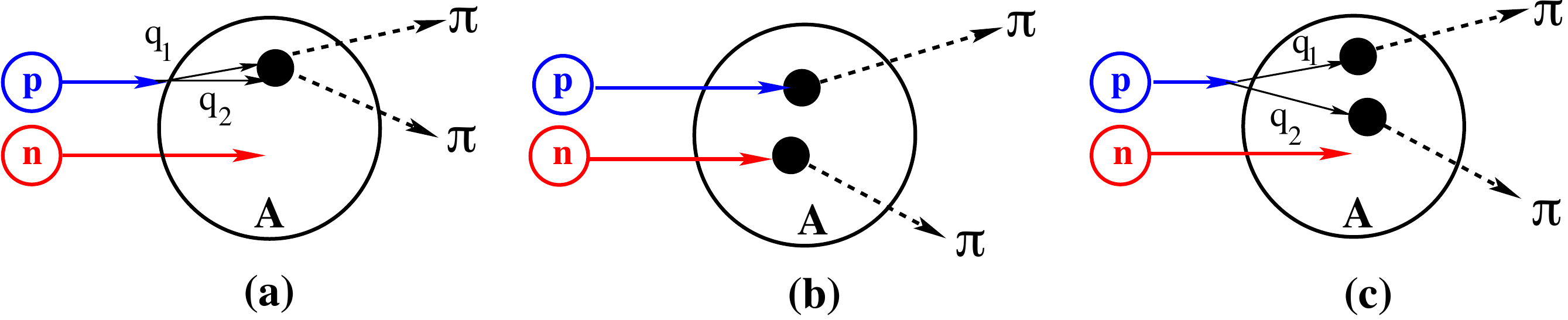} 
   \caption{Three  double parton mechanisms of dipion production.}
    \label{sketch}
 \end{figure}
Hence we conclude that the current experiments at RHIC have found a signal of double-parton interactions and that  future experiments at RHIC will be able to obtain a unique information about  double quark distributions in nucleons. It will be crucial for such studies to perform analyses for smaller bins in $\eta$ and preferably switch to the  analysis in bins of Feynman $x$.

 \section{Production of two forward pions and double-parton mechanism in $dAu$ scattering}
Let us extend now apply our results to the case of d-A scattering studied at RHIC.
In this case there are three distinctive double-parton mechanisms depicted in Fig.~ 
\ref{sketch}. The first two are the same as in the $pA$ scattering - scattering of two partons of the nucleon off two partons belonging to different nucleons (mechanism a), and off two   partons belonging to the same nucleon of the target (mechanism b) \cite{Strikman:2001gz}. The third mechanism, which is not present for $pA$ scattering is scattering of one parton of proton and one parton of the neutron off two partons of the nucleus.
Let us consider the ratio of the double-parton and leading twist contributions for $dA$ and $pp$ collisions 
\begin{equation}
r_{dA}= r_a+r_b+r_c= {\sigma_{DP}(dA)\over \sigma_{LT}(dA)}/{\sigma_{DP}(pp)\over \sigma_{LT}(pp)}.
\end{equation}
The contribution to $r_A$ of the mechanisms (a), (c) is given by  
\cite{Strikman:2001gz}:
\begin{equation}
r_c= T(b)\sigma_{eff}; r_a = 1,
\end{equation}
where $T(b)$ is the standard nuclear profile function ($\int d^2bT(b)=A$).
Here we neglected nuclear gluon shadowing effect which is a small correction for the double-parton mechanism (cf. Ref.\cite{GSV}) but maybe important for the LT mechanism where $x_g $ maybe as low as $10^{-3}$ due to    the leading twist shadowing (see discussion below).

\begin{figure}[t]  
   \centering
   \includegraphics[width=0.9\textwidth]{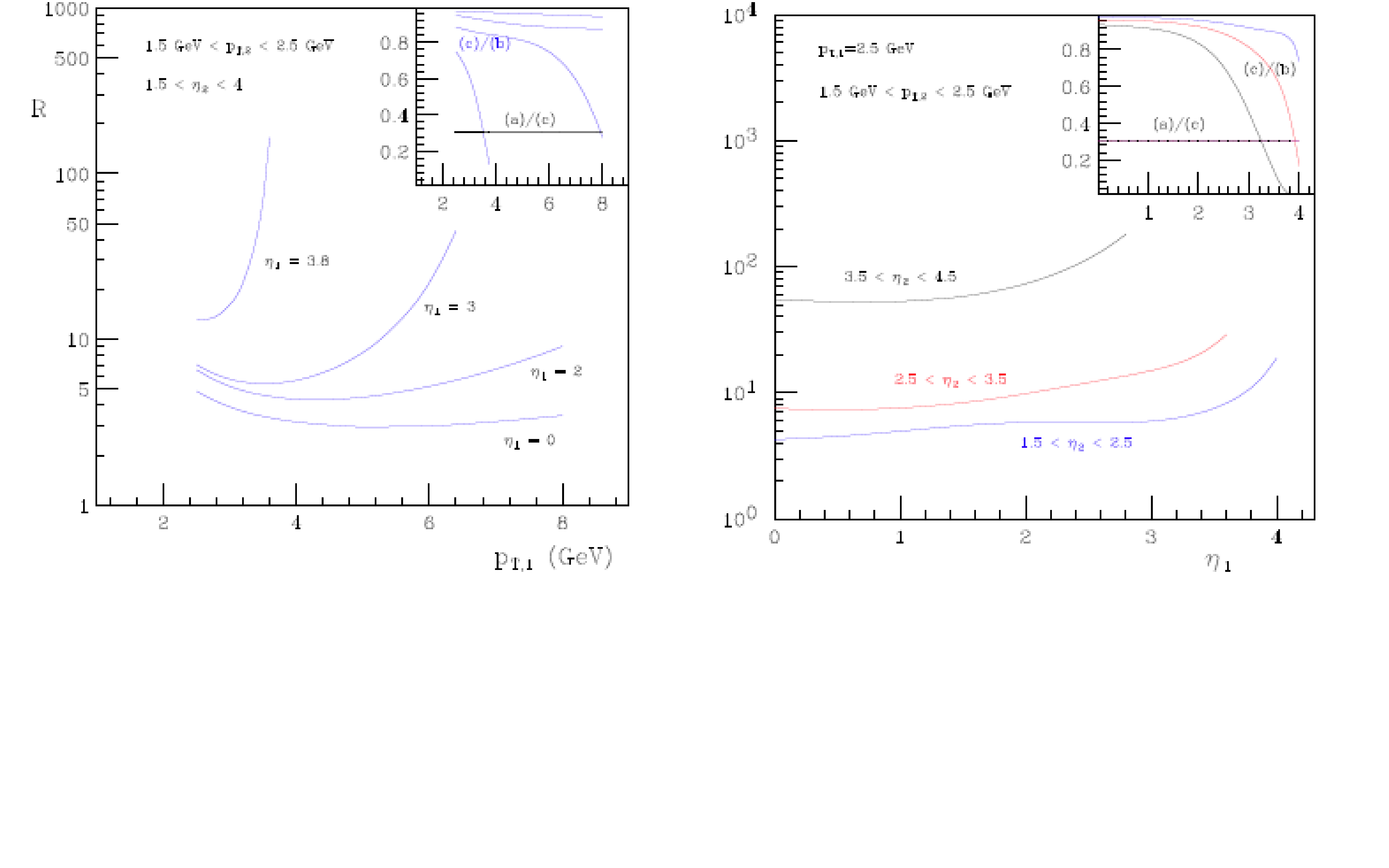}
   \vspace*{-25mm} 
   \caption{Ratio of double-parton leading-twist contributions
in $dA\to\pi^0\pi^0X$.
The inserts show the ratios $r_c/r_b$ and $r_a/r_c$.}
    \label{doubleda}
 \end{figure}

 For the central d-Au collisions $T_A\approx 2.2 fm^{-2}$  and so $r_b/r_c\sim 1/3$ and $r_b+r_c \sim 4.4$. 
The contribution (b) can be  calculated in a model independent way since  no parton correlations enter in this case. $r_c\approx r_b$ for moderate rapidities where 
correlations between partons are not important. It reaches $ r_c \sim 2r_b/3$ for the kinematics where only valence quarks contribute but $x_q+ x_{q'} \ll 1$.
For the very forward region $r_c\ll r_b$ since the  kinematic constrain $x_q+ x_{q'} \le 1$ is not present in this case. 
As a result $r_{dAu}$ for small $b$ becomes of the order ten, see Fig.~ \ref{doubleda} ($r_A$ changes from $\sim$  9 to $\sim $ 12 for
$\pi R_{int}^2 = 15 \div 20$ mb. 
 
 Since the single inclusive pion spectrum for $\eta_2 \sim 2 \div 3$ is suppressed by a factor of the order $R_A(b)= 1/3 \div 1/4$ we find for the ratio of the pedestals in $dAu$ and $pp$:
 
\begin{equation}
R_{pedestal}= r_A R_A(b) \sim 2.5  \div 4,
\end{equation}
which should be compared with the experimental value of $R_{pedestal}\sim 3$. 
Hence we naturally explain the magnitude of the enhancement of the pedestal in central $dAu$ collision.

If  most of the pedestal in the kinematics studied at RHIC is due to the double-parton mechanism,  the uncertainties in the estimate of the rates due to this mechanism   and uncertainties in the strength of the suppression of the single inclusive forward pion spectrum at $b\sim 0$ would make it  very difficult to subtract this contribution with a precision necessary to find out whether all pedestal is due to double-parton  mechanism or there is a room for  a small contribution of the $2\to 1$ broadening mechanism  as it was assumed  in \cite{Albacete:2010pg}.

 
 The suppression of the away peak originating from the LT contribution is  due to two effects: (i) the gluon shadowing 
  for $x\sim 10^{-3}$ and 
$b\le  3\, fm$ and $Q^2 \sim \mbox{few GeV}^2$ reduces the cross section by a factor of about two 
(cf. Fig.~1), (ii) stronger effect of effective fractional energy losses due to larger $x$ of the quark in the LT mechanism than in the double parton mechanism, leading to a suppression factor of the order two \cite{SV}. Combined these effects result in  a suppression of the order of four as compared to the single pion trigger, and overall suppression of the order of ten.  This is pretty close to the maximal possible 
 suppression which could be estimated as 
the probability of 
  the "punch through" mechanism - contribution from the process where a quark scatters off one nucleon but does not encounter any extra nucleons at its impact parameter. Probability of such collisions at $b\sim 0$ for interaction with Au nucleus is of the order 
$5 \div 10\%$ \cite{Alvioli:2009ab}. 

The preliminary  data \cite{starqm09} are described well as a sum of the  away peak of the strength $\le 1/4$ of the strength observed in $pp$ scattering,   and the pedestal enhanced by a factor of three
relative to the $pp$  case,   corresponding to 
 reduction of the away peak relative to pedestal of the order of ten.

 .

 \section{Conclusions}
   
   The LT twist nuclear shadowing can be estimated with a small uncertainty using correspondence with hard diffraction leading to the expectation of a large gluon shadowing  at $x\le 10^{-3}$. This influences significantly estimates of the proximity to the black disk regime, so that for small $b$ they are reached for proton and nuclei at comparable energies. For $b\sim 0 $ black disk regime is effective for the RHIC energies leading to post selection effect (effective fractional energy losses) for the propagation of the leading partons. This is the only effect which currently allows to explain the suppression of the $2\to 2$ mechanism for production of leading pions which dominates in the $pp$ scattering including survival of forward - central correlations and suppression of forward - forward correlations. The processes of the production of two forward pions in $pp, dAu$ scattering due to double -parton mechanism explain the pedestal observed at RHIC  and make it difficult to search for the signal of $2\to 1$ processes. Magnitude of the away peak is consistent with the expected magnitude of the suppression of the $2\to 2$  mechanism due to LT gluon shadowing and fractional energy losses due to the post-selection mechanism.
   
   Also these experiments for the first time provide an opportunity to observe correlations of leading quarks in protons. 

\section*{Acknowledgments}
I am  grateful to  L.~Frankfurt, B.~Jacak, and L.~McLerran for 
useful discussions and comments. My special thanks are to Werner Vogelsang for joint work on the double parton interaction  mechanism of the two pion production (section 5 of the contribution).   I  would like to thank also the Yukawa International
Program for Quark--Hadron Sciences  for the hospitality  and  the stimulating atmosphere during 
a part  of this study.   The research was supported 
by DOE grant No. DE-FG02-93ER40771. 


\bibliographystyle{elsarticle-num}
\bibliography{<your-bib-database>}

\begin{thebibliography}{00}
\bibitem{Gribov:1968jf}
  V.~N.~Gribov,
  Sov.\ Phys.\ JETP {\bf 29} (1969) 483
  [Zh.\ Eksp.\ Teor.\ Fiz.\  {\bf 56} (1969) 892].
   \bibitem{Abramovsky:1973fm}
  V.~A.~Abramovsky, V.~N.~Gribov and O.~V.~Kancheli,
  Yad.\ Fiz.\  {\bf 18} (1973) 595
  [Sov.\ J.\ Nucl.\ Phys.\  {\bf 18} (1974) 308]
  
  
  \bibitem{Frankfurt:1998ym}
  L.~Frankfurt and M.~Strikman,
  Eur.\ Phys.\ J.\  A {\bf 5} (1999) 293;
  L.~Frankfurt, V.~Guzey and M.~Strikman,
  Phys.\ Rev.\  D {\bf 71} (2005) 054001; for a detailed discussion see 
  L.~Frankfurt, V.~Guzey and M.~Strikman, to be submitted to Phys.Rep.
\bibitem{Collins:1997sr}
  J.~C.~Collins,
  Phys.\ Rev.\  D {\bf 57} (1998) 3051
  [Erratum-ibid.\  D {\bf 61} (2000) 019902]
  \bibitem{Guzey:2009jr}
  V.~Guzey and M.~Strikman,
  Phys.\ Lett.\  B {\bf 687} (2010) 167
  [arXiv:0908.1149 [hep-ph]].

  \bibitem{Frankfurt:1996ri}
  L.~Frankfurt, A.~Radyushkin and M.~Strikman,
  Phys.\ Rev.\  D {\bf 55} (1997) 98

\bibitem{Rogers:2003vi}
  T.~Rogers, V.~Guzey, M.~Strikman and X.~Zu,
  Phys.\ Rev.\  D {\bf 69} (2004) 074011 

\bibitem{Frankfurt:2001nt}
  L.~Frankfurt, V.~Guzey, M.~McDermott and M.~Strikman,
  Phys.\ Rev.\ Lett.\  {\bf 87}, 192301 (2001)
\bibitem{Frankfurt:2007rn}
   L.~Frankfurt and M.~Strikman,
  Phys.\ Lett.\  B {\bf 645} (2007) 412
\bibitem{Rakness} G.~Rakness, private communication


\bibitem{Werner} F.~Aversa, P.~Chiappetta, M.~Greco and J.~P.~Guillet,
  Nucl.\ Phys.\ B {\bf 327}, 105 (1989);  B.~Jager, A.~Schafer, M.~Stratmann and W.~Vogelsang,
  Phys.\ Rev.\ D {\bf 67},  (2003) 054005;
  D.~de Florian,
  Phys.\ Rev.\ D {\bf 67},  (2003) 054004 
\bibitem{GSV}
           V.~Guzey, M.~Strikman, and W.~Vogelsang,
           Phys.\ Lett.\ B {\bf 603}, (2004) 173 . 
\bibitem{star}
 J.~Adams {\it et al.}  [STAR Collaboration],
  Phys.\ Rev.\ Lett.\  {\bf 97}, (2006) 152302.
  \bibitem{phenix2}B. Meredith, Nucl.\  Phys. \ {\bf A830 } (2009) 595 [nucl-ex/0907.4832].
  \bibitem{Kharzeev}  D.~Kharzeev, E.~Levin and L.~McLerran,
  Phys.\ Lett.\ B {\bf 561}, (2003) 93;
 D.~Kharzeev, Y.~V.~Kovchegov and K.~Tuchin,
  Phys.\ Rev.\ D {\bf 68}, (2003) 094013.

  \bibitem{Dumitru:2005gt}
  A.~Dumitru, A.~Hayashigaki and J.~Jalilian-Marian,
  Nucl.\ Phys.\  A {\bf 765} (2006) 464
\bibitem{starqm09} E. Braidot for the STAR collaboration, arXiv:1005.2378.
\bibitem{phenixqm09} B. Meredith, DIS 2010 Conference Proceedings, PoS(DIS 2010)081.
\bibitem{SV}   M.~Strikman and W.~Vogelsang,
  arXiv:1009.6123 [hep-ph].
\bibitem{BDFS} B.~Blok, Yu.~Dokshitzer, L.~Frankfurt and M.~Strikman,
  arXiv:1009.2714 [hep-ph].
\bibitem{Frankfurt:2003td}
  L.~Frankfurt, M.~Strikman and C.~Weiss,
  Phys.\ Rev.\  D {\bf 69} (2004) 114010
  [arXiv:hep-ph/0311231].

\bibitem{Strikman:2001gz}
  M.~Strikman and D.~Treleani,
  Phys.\ Rev.\ Lett.\  {\bf 88} (2002) 031801 
  [arXiv:hep-ph/0111468].
\bibitem{Albacete:2010pg}
  J.~L.~Albacete and C.~Marquet,
  arXiv:1005.4065 [hep-ph].

\bibitem{Alvioli:2009ab}
  M.~Alvioli, H.~J.~Drescher and M.~Strikman,
  Phys.\ Lett.\  B {\bf 680}, 225 (2009)
  [arXiv:0905.2670 [nucl-th]], and to be published.
 \end{thebibliography}



\end{document}